\documentclass{aa}
\usepackage{graphicx}
\usepackage{longtable}
\usepackage{supertabular}

\def\arcsec{\tt ''}
\def\simgt{\ {\raise-.5ex\hbox{$\buildrel>\over\sim$}}\ }
\def\simlt{\ {\raise-.5ex\hbox{$\buildrel<\over\sim$}}\ }

\begin{document}

\title{$uvby\beta$ photometry of early type open cluster and 
field stars\thanks{Based on measurements obtained at McDonald Observatory 
of the University of Texas at Austin}$^,$\thanks{Tables 3 - 6 are only electronically available via the CDS}}
\authorrunning{G. Handler}
\titlerunning{$uvby\beta$ photometry of early type stars
}
   \author{G. Handler
          }
   \offprints{G. Handler}

   \institute{
Institut f\"ur Astronomie, Universit\"at Wien, T\"urkenschanzstra\ss e 17,
1180 Wien, Austria (gerald.handler@univie.ac.at)}

\date{Received January 14, 2011; Accepted February 2, 2011}

\abstract
% context heading (optional)
{The $\beta$ Cephei stars and slowly pulsating B (SPB) stars are massive 
main sequence variables. The strength of their pulsational driving 
strongly depends on the opacity of iron-group elements. As many of those 
stars naturally occur in young open clusters, whose metallicities can be 
determined in several fundamental ways, it is logical to study the 
incidence of pulsation in several young open clusters.}
% aims heading (mandatory)
{To provide the foundation for such an investigation, Str\"omgren-Crawford
$uvby\beta$ photometry of open cluster target stars was carried out to 
determine effective temperatures, luminosities, and therefore cluster 
memberships.}
% methods heading (mandatory)
{In the course of three observing runs, $uvby\beta$ photometry for 168
target stars was acquired and transformed into the standard system by 
measurements of 117 standard stars. The list of target stars also 
included some known cluster and field $\beta$ Cephei stars, as well as 
$\beta$ Cephei and SPB candidates that are targets of the asteroseismic 
part of the Kepler satellite mission.}
% results heading (mandatory)
{The $uvby\beta$ photometric results are presented. The data are shown to 
be on the standard system, and the properties of the target stars are 
discussed: 140 of these are indeed OB stars, a total of 101 targets
lie within the $\beta$~Cephei and/or SPB star instability strips, and each 
investigated cluster contains such potential pulsators.}
% conclusions heading (optional)
{These measurements will be taken advantage of in a number of subsequent 
publications.} 


\maketitle

\section{Introduction}

The $\beta$ Cephei stars are a group of pulsating main sequence 
variables with early B spectral types. They oscillate in radial and 
nonradial pressure modes with typical periods of several hours. Stankov 
\& Handler (\cite{SH05}) provide an overview of those stars. Pigulski \& 
Pojma{\'n}ski (\cite{PP08}) doubled the number of class members to about 
200. As these are young massive stars (and are thus progenitors of type 
II supernovae), they naturally occur in the galactic plane, in open 
clusters, and stellar associations. In general, this statement also 
holds for the less massive slowly pulsating B (SPB) stars. They 
neighbour the $\beta$ Cephei stars in the HR diagram, but they are 
cooler and less luminous, and they pulsate in gravity modes with periods 
of a few days (see, e.g., De Cat \cite{PDC07}).

The physical origin of pulsation driving of the $\beta$ Cephei and SPB 
stars is well established (Moskalik \& Dziembowski \cite{MD92}), and is 
caused by the huge number of transitions inside the thin structure of 
the electron shells in excited ions of the iron-group elements (Rogers 
\& Iglesias \cite{RI94}): the $\kappa$ mechanism. Obviously, the power 
of pulsational driving will strongly depend on the abundance of 
iron-group elements and on their opacities in the driving zone. Credible 
pulsational models must reflect the conditions inside the real stars, 
reproducing all observables such as the extents of the $\beta$ Cephei 
and SPB instability strips and their metallicity dependence. These 
depend on the input data used in the models, which can therefore be 
tested.

The metallicities of stellar aggregates can be determined in several 
fundamental ways. The incidence of core hydrogen-burning B-type 
pulsators among open cluster stars can then yield important constraints 
on what abundance of metals (and, by extrapolation, amount of iron group 
elements) is required to drive their oscillations. The aim of the 
present and subsequent works is to determine observationally the 
incidence of $\beta$ Cephei and SPB stars in a number of open clusters 
exactly for this purpose.

Ground-based measurements of stellar variability are hampered by the
presence of the Earth's contaminated atmosphere. Scintillation and
variable transparency of the night sky limit the precision of stellar
brightness measurements. Therefore, the level at which the presence of 
oscillations in a given star can be detected is finite. Although there are 
techniques that optimize the precision of ground-based photometric 
measurements (again, often taking advantage of stellar clusters), 
observations from space are superior given a large enough telescope.

The {\it Kepler} mission, the most powerful instrument for measuring 
stellar brightness variations to date (Koch et al. \cite{KBB10}), aims 
at detecting transits of extrasolar planets in the habitable zone around 
their host stars. As the only inhabited planet known so far revolves 
around a middle-aged main sequence G star, the sample of target stars of 
the {\it Kepler} mission was chosen to observe as many similar stars as 
possible to the highest precision. Stars at such an age do not dominate 
the population at low galactic latitudes, so the {\it Kepler} field was 
chosen to be some 10\degr off the galactic plane (Batalha et al. 
\cite{BBK10} and references therein). $\beta$~Cephei and SPB stars with 
magnitudes of $V>7.5$ formed in the galactic plane would hardly reach 
these galactic latitudes within their main sequence life times and are 
thus expected to be unusual. Therefore, characterizing {\it Kepler} 
$\beta$~Cephei and SPB star candidates is important.

The present paper reports the results of a study of bona fide and 
candidate field and open cluster $\beta$ Cephei stars in the Str\"omgren 
photometric system: 168 target stars were measured, 107 of them being 
open cluster stars, 17 known cluster and field $\beta$ Cephei stars, and 
42 Kepler targets. To transform the data into the standard system, 117 
Str\"omgren photometric standards were measured as well. The outcome of 
this study will be used in subsequent papers.

\section{Observations}

\subsection {Measurements and reductions}

The measurements were obtained with the 2.1-m telescope at McDonald 
Observatory in Texas. Three observing runs were carried out in October 
2008, March 2009, and August/September 2010. The first two observing 
runs were dedicated to stars in open clusters and known field 
$\beta$~Cephei stars, whereas the third run focused on Kepler targets 
and on supplementary $H_\beta$ measurements of previous targets missing 
this information.

In all runs, a two-channel photoelectric photometer was used, but only 
employed channel 1. The same filter set, the same photomultiplier tube, 
and the same operating voltage were used during all observations. The 
only variables in the observational setup were the reflectivities of the 
telescope's mirrors: the primary mirror was not cleaned or aluminized 
within the time span of the observing runs, but dust on the secondary 
mirror is blown off on a monthly basis.

Photometric apertures of 14.5 and $29\arcsec$ were used in most cases, 
depending on the brightness of the target and sky background as well as 
on crowding of the field. In a few cases of extreme crowding or of a 
close companion, a $11\arcsec$ aperture and extremely careful (offset) 
guiding had to be used. As the photometer's filter wheel can only carry 
four filters at once, the $uvby$ measurements had to be taken separately 
from the H$_{\beta}$ data. No H$_{\beta}$ measurements were taken for 
open cluster targets that were immediately identified as non-OB stars 
from their Str\"omgren "bracket quantities" (see Sect.\ 4 for details).

As the measurements aimed at obtaining estimates of the effective 
temperatures and luminosities of most targets possible rather than 
establishing new standard stars, most stars were observed only once. A 
few exceptions were made for standard and target stars that were used 
for purposes of determining extinction coefficients, for target stars 
that were deemed the most interesting astrophysically, or where a 
previous measurement appeared suspicious.

\subsection {Selection of standard stars}

A set of standard stars was selected to span the whole parameter range 
of the targets in terms of $(b-y)$, $m_1$, $c_1$, $\beta$, and $E(b-y)$. 
It was observed for transforming the measurements into the standard 
system. For reasons of homogeneity in the colour transformations, the 
majority of the adopted standard Str\"omgren indices were taken from the 
work of a single group of researchers. The standard stars were chosen 
from the papers on NGC 1502 (Crawford \cite{Cr94}), IC~4665 (Crawford \& 
Barnes \cite{CB72}), NGC 2169 (Perry, Lee, \& Barnes \cite{PLB78}), NGC 
6910 and NGC 6913 (Crawford, Barnes, \& Hill \cite{CBH77}), O-type stars 
(Crawford \cite{Cr75}), h and $\chi$ Per (Crawford, Glaspey, \& Perry 
\cite{CGP70}), Cep OB3 (Crawford \& Barnes \cite{CB70}), Lac OB1 
(Crawford \& Warren \cite{CW76}), and on three field stars (Crawford et 
al. \cite{CBG72}, Knude \cite{JK77}).

\subsection {Data reduction}

The data were reduced in a standard way. The instrumental system's 
deadtime of 33 ns was determined by measuring the twilight sky, and then 
was used to correct for coincidence losses. Sky background subtraction 
was done next, followed by nightly extinction corrections determined 
from measurements of extinction stars that also served as standards. The 
applied extinction coefficients varied between $0.14 - 0.18$ in $y$, 
$0.054 - 0.069$ in $(b-y)$, $0.050 - 0.064$ in $m_1$, and $0.126 - 
0.157$ in $c_1$.

%demonstrating the homogeneous quality of the usable observing nights. 
%Ideally, one would want to determine extinction coefficients 
%simultaneously with the transformation equations (Sterken \& Manfroid 
%reference). However, as some correlations in the coefficients of the 
%transformation equations and variations in the instrumental system were 
%present, a less precise, but hopefully more accurate, step-by-step 
%procedure was adopted. As verified later, the maximum possible 
%systematic error introduced by this approach was generally about an 
%order of magnitude smaller than the rms scatter of an individual 
%measurement.

\section{Transformation equations}

The equation for $(b-y)$ only has two parameters, so we only needed to
calculate a linear fit to the data. However, as it turned out, the
photometric zeropoints of the three individual observing runs and seasons 
were different (most likely as a consequence of the large temporal gaps 
between the observing runs) and had to be determined separately. After 
adjustment of the zeropoints, the slope of the transformation was 
re-determined, and the procedure repeated until convergence. The final 
transformation equation was
\begin{equation}
(b-y)=1.0563 (b-y)_N + zpt(b-y),
\end{equation}
where the subscript $N$ denotes the colour in the natural system, and
$zpt(b-y)$ is the zeropoint of the transformation equation, listed in
Table 1. The rms residual scatter of a single standard star measurement in
$(b-y)$ is an unsatisfactory 13.4 mmag.

\begin{table}
\caption{$(b-y)$ colour transformation zeropoints}
\begin{tabular}{llc}
\hline
Observing run & Standard stars & $zpt (b-y)$\\
\hline
Autumn 2008 & O stars & 1.3497 $\pm$ 0.0026\\
Autumn 2008 & Cep OB3 & 1.3514 $\pm$ 0.0034\\
Autumn 2008 & h \& $\chi$ Per & 1.3413 $\pm$ 0.0025\\
Autumn 2008 & NGC 6910/13 & 1.3566 $\pm$ 0.0021\\
\hline
Autumn 2008 & above combined & 1.3485 $\pm$ 0.0014\\
Spring 2009 & NGC 1502, 2169, 2244 & 1.3916 $\pm$ 0.0037\\
Autumn 2010 & Lac OB1, field & 1.3302 $\pm$ 0.0023\\
\hline 
\end{tabular}
\end{table}

However, this high residual scatter does not mean that the present 
measurements are imprecise. Some standard stars were measured more than 
once, which indicates the precision of the data. The average rms scatter 
of the $(b-y)$ values of standard stars that were measured three times 
is only 2.0 mmag.

It is worth noting that the $(b-y)$ transformation zeropoints are 
different by up to $6\sigma$ when standard stars from different 
publications are considered (upper part of Table~1). This comparison 
only uses data from the most fruitful observing run in Autumn 2008, 
where several of the different groups of standard stars were measured in 
the same nights. The total 13.4 mmag residual scatter in $(b-y)$ may 
therefore be due to a combination of underestimation of the precision of 
the data and of imperfections in the standard values adopted.

Because accuracy is more important than precision (see, e.g., Bevington 
\cite{B69} for the distinction between these two terms) in the present 
case, the same transformation slope was used for all $(b-y)$ 
measurements, but seasonal (lower part of Table 1) zeropoints were 
applied. In other words, it is assumed that the changes in the seasonal 
zeropoints of the colour equations are dominated by variations in the 
instrumental system.

The remaining transformation equations are to be determined by a 
(simultaneous) three-parameter fit to the measurements of the standard 
stars as a colour correction by means of the $(b-y)$ data is necessary. 
The equation for $m_1$ derived by simultaneously fitting three 
parameters appears biased from correlations in the $m_1$ and $b-y$ 
indices due to reddening: $E(m_1)=-0.32E(b-y)$. The range spanned by the 
$m_1$ values of the standard stars is 0.37 mag, the range in $(b-y)$ is 
0.89 mag, 2.4 times larger.

Therefore, the measured and standard $m_1$ values were linearly
fitted first, and only then were the $(b-y)$ correction term and the 
zeropoint fixed. This resulted in the following transformation equation
\begin{equation}
m_1=1.0195 m_{1,N}-0.0162 (b-y)_N-0.8469.
\end{equation}
Statistically insignificant variations occurred in the zeropoint when
different ensembles of comparison stars were considered. The rms residual
of a single standard $m_1$ measurement is 12.1 mmag.

%\begin{table}
%\caption{Seasonal $m_1$ and $c_1$ index transformation zeropoints}
%\begin{tabular}{lcc}
%\hline
%Season & $zpt(m_1)$ & $zpt(c_1)$\\
%\hline
%Autumn 2008 & $-0.8436 \pm$ 0.0012 & $-0.5740 \pm 0.0016$\\
%Spring 2009 & $-0.8568 \pm$ 0.0021 & $-0.5693 \pm 0.0031$\\
%Autumn 2010 & $-0.8507 \pm$ 0.0022 & $-0.5961 \pm 0.0020$\\
%\hline 
%\end{tabular}
%\end{table}

Concerning $c_1$, correlations between the coefficients in the 
transformation equation due to reddening are also to be expected, but 
are less severe than in $m_1$ because the $c_1$ values have a much 
wider spread than $m_1$ and because $c_1$ is less affected by reddening 
than $m_1$. A simultaneous three-parameter linear fit yielded
\begin{equation}
c_1=1.0025 c_{1,N}+0.1018 (b-y)_N-0.5484.
\end{equation}
The seasonal zeropoints were roughly, but not fully satisfactorily, 
consistent. Again, as accuracy is more important than precision, a 
single zeropoint was adopted for all data sets. The residual scatter of 
the standard star measurements transformed in this way is 15.6 mmag per 
single point.

No difficulties with varying zeropoints were encountered when determining
the transformation equation for the $\beta$ value. This is no surprise as
it is a differential measurement at the same effective wavelength. The
transformation equation for $\beta$ is
\begin{equation}
\beta=0.8302\beta_N-0.0439(b-y)_N+0.9532,
\end{equation}
leaving a residual scatter of $11.4$~mmag per single measurement.  
Finally, the transformation equations for the $V$ magnitude require
nightly zeropoints (Table 2) to take variable sky transparency into
account. As some papers reporting standard Str\"omgren colour indices do
not quote $V$ magnitudes, these values were supplemented by literature
data as supplied by the SIMBAD data base and cross-checked with the
original references. The final transformation was

\begin{table}
\caption{Nightly $V$ magnitude transformation zeropoints}
\begin{tabular}{lc}
\hline
Civil date & $zpt(y)$\\
\hline
07 Oct 2008 & $20.022 \pm 0.004$\\
08 Oct 2008 & $20.036 \pm 0.006$\\
09 Oct 2008 & $20.001 \pm 0.007$\\
16 Oct 2008 & $20.003 \pm 0.004$\\
04 Mar 2009 & $20.102 \pm 0.005$\\
01 Oct 2010 & $19.739 \pm 0.005$\\
02 Oct 2010 & $19.703 \pm 0.008$\\
\hline
\end{tabular}
\end{table}

\begin{equation}
V=0.9961y_N+0.0425(b-y)_N+zpt(y),
\end{equation}
resulting in a residual scatter of $22.2$~mmag per single measurement. 
Observations yielding statistically significant outliers in each of the 
transformation equations were excluded from the determination of its 
parameters and are marked as such in the data tables that follow. It 
cannot be judged whether this indicates a problem with the present 
measurements or with the standard values used.

\section{Results}

With the transformation equations in place, the colour indices in the 
standard system can be determined for all standard and target stars. The 
results are listed in Tables 3 - 6. Table 3 contains the present 
measurements of the standard stars themselves, transformed into the 
standard system. Table 4 reports the $uvby\beta$ photometry for open 
cluster target stars not previously known to pulsate. Table 5 lists the 
Str\"omgren-Crawford photometry for known $\beta$~Cephei stars plus a 
few other targets. Finally, Table 6 contains the results for stars in 
the {\it Kepler} field.

In the following, stars in open clusters are always designated with the 
cluster name followed by their identification in the WEBDA\footnote{\tt 
http://www.univie.ac.at/webda/} data base. Measurements of standard 
stars that were rejected for computing the transformation equations (or 
where no $V$ magnitudes or $H_{\beta}$ values were available in the 
literature) were treated in the same way as target star observations and 
are marked with asterisks in Table 3.

Some of the stars used as standards have been shown to be intrinsically 
variable in the literature. However, standard stars must have 
temperatures and luminosities similar to the targets that would ideally 
be pulsating variables. Therefore the use of variable standard stars of 
low amplitude cannot be avoided. Intrinsic variability of standard stars 
not exceeding the accuracy of the present data is therefore tolerable 
and measurements that are significantly off the limits would be rejected 
anyway.

\subsection{Comments on individual stars}

BD+36 4867 was mistakenly observed when intending to measure the 
$uvby\beta$ standard star BD+36 4868. This error came from confusion of 
the coordinates of the two stars in the SIMBAD data base at the time of 
the measurements. The Str\"omgren indices of BD+36 4867 are listed for 
completeness in Table 6, indicating a mid G-type star.

The published $V$ magnitudes of NGC 1893 196 vary between 12.30 and 
12.79. This unusually wide range raises the suspicion of stellar 
variability. Table 4 lists $V=12.637$ and $\beta=2.441$. The latter 
value indicates strong hydrogen-line emission, as demonstrated 
spectroscopically (Marco et al. \cite{MBN01}).

\addtocounter{table}{4}

\section{Analysis and discussion}

\subsection{Validity of the transformation equations}

The ranges in which the transformation equations are valid are examined in
Fig.\ 1. It shows the distributions of the standard and target star
measurements with respect to the different $uvby\beta$ colour indices and
reddening. The routines by Napiwotzki, Sch\"onberner, \& Wenske
(\cite{NSW93}) were used to derive the latter.

\begin{figure}
\includegraphics[width=80mm,viewport=-10 00 255 728]{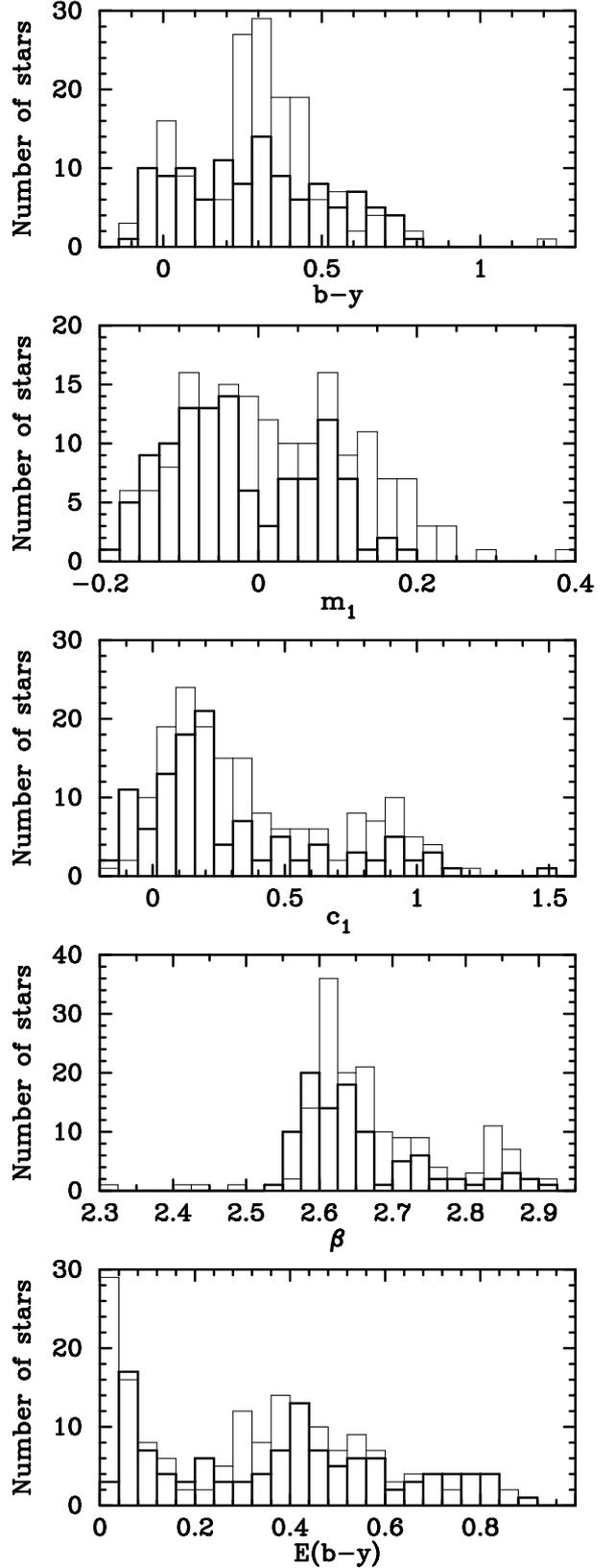}
\caption{Distributions of the different Str\"omgren-Crawford colour 
indices amongst standard (thick histogram bars) and target (thin 
histogram bars) stars.}
\end{figure}

The $(b-y)$ values of all but one target star (Roslund 2 13, a very red 
object) are contained within the range spanned by the standard stars. 
The same comment is true for the $c_1$ parameter and reddening $E(b-y)$. 
Sixteen (i.e.\ 10\%) of the targets have more positive $m_1$ values than 
any standard star. These are stars of later spectral types than A0 which 
are not the prime interest of this work.

As far as $H_\beta$ is concerned, five stars with values below 2.55 were 
observed, including two (supposed) standard and three target stars. Both 
standard stars were rejected after determining the transformation 
equations due to high residual deviations. It is suspected that all five 
of these stars are Be stars. The hydrogen line emission of such stars is 
often variable (e.g., McSwain, Huang, \& Gies \cite{MHG09}) which 
explains the high residuals and makes the tabulated values unreliable. 
They are listed for completeness only.

Considering the distribution in $E(b-y)$, about two thirds of the stars 
with the smallest reddening are among the {\it Kepler} targets: the 
satellite's field of view deliberately excludes the central galactic 
plane. Two of the remaining targets are $\beta$ Cephei stars of rather 
high galactic latitude, and the remainder are cool main sequence stars 
in the foreground of some of the target open clusters. In Tables 4 - 6 
the colour indices that are outside the range of those spanned by the 
standard stars are marked with colons and should be used with caution.

\subsection{Are the present data on the standard system?}

Before inferring physical parameters of the targets, it must be made sure
that the data are commensurate with the standard system. It is a subtle
process to obtain accurate standard photometry of reddened early-type
stars, see, e.g., Crawford (\cite{Cr99}) for a discussion.

One test is to compare published $(U-B)$ colours with $(u-b)$ values 
from Str\"omgren indices (see Crawford \cite{Cr94}) and to compare the 
resulting relation with the one defined by standard stars. This is done 
in Fig.\ 2, using the results for target stars with existing UBV 
photometry. The $(U-B)$ values for the target stars were taken from the 
General Catalogue of Photometric Data (Mermilliod, Mermilliod, \& Hauck 
\cite{MMH97}).

For easier visual inspection, the slope of the $(U-B)$ vs.\ $(u-b)$ 
relation was removed by a linear fit. The residuals are compared with 
those of the standard values for reddened O-type stars (Crawford 
\cite{Cr75}) and for bright stars earlier than B5 (Crawford, Barnes, \& 
Golson \cite{CBG71}), which are on the average considerably less 
reddened than the O stars. For better illustration, we only show a fit 
to the relations defined by the standard stars for comparison with the 
data of the target stars.

\clearpage

\begin{figure}
\includegraphics[width=80mm,viewport=00 00 270 255]{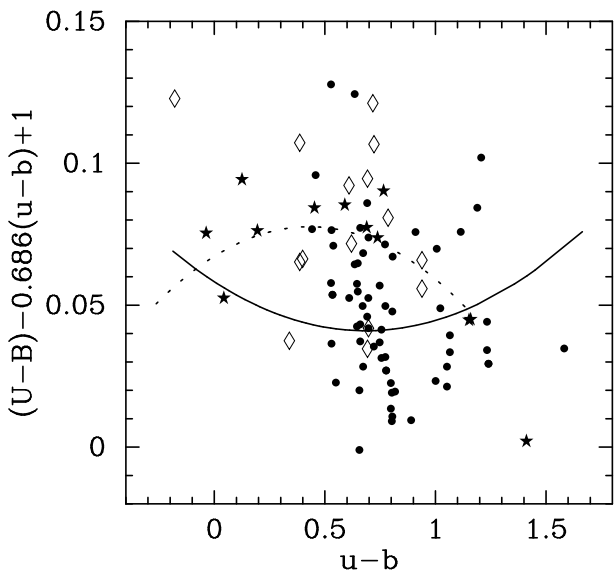}
\caption{Comparison of the present measurements and published Johnson 
photometry. Circles are open cluster targets not yet known to pulsate, 
diamonds are known $\beta$~Cephei stars with new Str\"omgren colour 
indices, and star symbols are early-type targets in the {\it Kepler} 
field. The dotted line is the relation defined by unreddened B stars, 
whereas the full line is the relation inferred for reddened OB stars. 
See text for more information.}
\end{figure}

The fits for the O and B-type stars in Fig.\ 2 are somewhat different.
However, the relation for the more strongly reddened target stars are not
systematically different from the one defined by the reddened O-type
standards, and the relation for the less reddened targets shows no
systematic offset from the one defined by the mildly reddened B-type
standards. The present $uvby\beta$ photometry is therefore on the standard
system.

\subsection{Distinguishing OB stars from cooler ones}

OB stars can be separated from objects of later spectral type by using 
the reddening independent Str\"omgren ``bracket quantities" 
$[m_1]=m_1+0.32(b-y)$ and $[c_1]=c_1-0.2(b-y)$. As a rule of thumb, 
stars with $[m_1]<0.14$ are B type stars and stars with $[m_1]>0.22$ are 
of spectral type A3 and later. Astrophysically, this separation is 
caused by the changing curvature of the stellar energy distribution 
depending on temperature. Figure 3 shows the distribution of the target 
and standard stars in an $[m_1],[c_1]$ diagram.

\begin{figure}
\includegraphics[width=80mm,viewport=00 00 270 270]{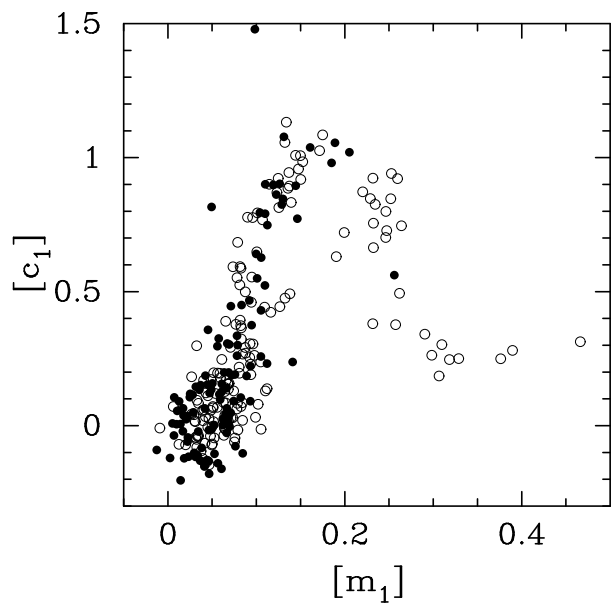}
\caption {Plot of the Str\"omgren "bracket quantities". These
reddening-free indices allow an easy separation between OB and cooler
stars; all objects with $[m_1]\simgt0.14$ are non-OB stars. One very cool
star lies outside the borders of this diagram. Filled circles are for
standard stars, open circles for the target stars.}
\end{figure}

All but one standard star were chosen to be of no later type than early 
A: 83\% of the targets are in the same domain. Of the 30 target stars 
that cannot be OB stars, twelve have been associated with the open 
cluster Berkeley 4, and should therefore be foreground stars. Seven 
non-OB stars are {\it Kepler} targets, and six were mentioned in 
connection with NGC 7380, therefore also not being cluster members.

\subsection{Effective temperatures and luminosities of the target stars}

The effective temperatures and absolute magnitudes of the target stars 
can be determined with the routines by Napiwotzki et al. (\cite{NSW93}, 
see their paper for accurate descriptions of the calibrations employed). 
Bolometric corrections by Flower (\cite{F96}) and a bolometric magnitude 
of $M_{\rm bol}=4.74$ for the Sun (Livingston \cite{L00}) were used to 
derive stellar luminosities. Figure 4 shows the targets' locations in a 
$\log T_{\rm eff} - \log L$ diagram, in comparison with theoretical 
pulsational instability strips (Zdravkov \& Pamyatnykh \cite{ZP08}).

\begin{figure*}
\includegraphics[width=120mm,viewport=-110 00 373 483]{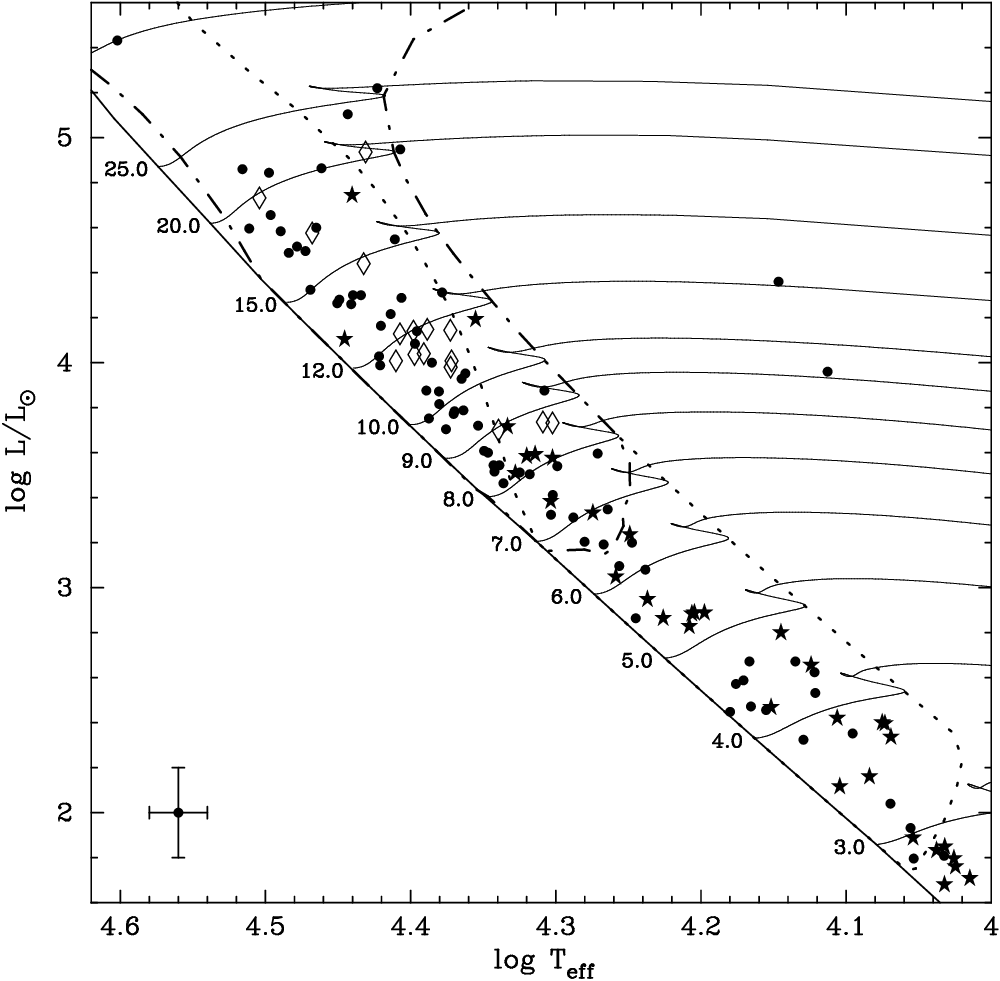}
\caption{$\log T_{\rm eff} - \log L$ diagram with the positions of 
the target stars, derived as explained in the text, indicated. 
Circles are open cluster targets not yet known to pulsate, diamonds are 
known $\beta$~Cephei stars with new Str\"omgren colour indices, and star 
symbols are early-type targets in the {\it Kepler} field. Some model 
evolutionary tracks are shown for comparison and are marked with the 
corresponding masses. The slanted full line is the zero age main 
sequence, the dotted line defines the theoretical SPB star instability 
strip, and the dashed-dotted line is the theoretical $\beta$~Cephei star 
instability strip.}
\end{figure*}

All targets previously known as $\beta$~Cephei stars are located within 
the corresponding instability strip. The catalogue of Galactic 
$\beta$~Cephei stars (Stankov \& Handler \cite{SH05}) only contains one 
object with a mass above 17 $M_{\sun}$, which could be a Be star, hence 
have overestimated luminosity from $H_{\beta}$ photometry. In contrast, 
the present, considerably smaller, sample contains three stars with 
$17.5<M/M_{\sun}<21$. There is one {\it Kepler} target in the high mass 
domain, which, however, appears to be a close binary with no pulsational 
light variation (Balona et al. \cite{BPD11}).

Table 7 summarizes how many of our target stars lie within the 
$\beta$~Cephei and SPB star instability strips, respectively, and how 
many are located in either and may therefore show both types of 
oscillations. Each of the open clusters observed contains potential 
pulsators and is therefore worthy of a variability search. As expected, 
the {\it Kepler} field contains only a few high-mass stars.

\begin{table}
\caption{Numbers of target stars within the $\beta$~Cephei or SPB 
star instability strip, or both}
\begin{tabular}{lccc}
\hline
Field & in $\beta$ Cep strip & in SPB strip & in both\\
\hline
ASCC 130 & 3/7 & 4/7 & 0/7\\
Berkeley 4 & 16/32 & 9/32 & 6/32\\
NGC 637 & 6/6 & 0/6 & 0/6\\
NGC 1893 & 10/12 & 2/12 & 1/12\\
NGC 2244 & 3/9 & 4/9 & 3/9\\
NGC 7380 & 12/27 & 9/27 & 3/27\\
Roslund 2 & 7/10 & 3/10 & 2/10\\
Kepler field & 10/42 & 26/42 & 8/42\\
\hline 
\end{tabular}
\end{table}

Two stars in Fig.\ 4 appear to be post main sequence objects. Berkeley 4 
513 is also known as LS I +63 98 and has been classified as an OBe star 
(Hardorp et al. \cite{HRS59}). The low $H_{\beta}$ value for the star 
supports this interpretation. A similar comment applies to NGC 7380 4 
that has been spectrally classified as B6Vne (Hoag \& Applequist 
\cite{HA65}). The post main sequence evolutionary status of these two 
stars may therefore just be apparent: the calibrations of $uvby\beta$ 
photometry are not applicable to emission line stars.

\section{Summary}

New $uvby\beta$ photometry was acquired for 168 open cluster and field 
stars, and was transformed into the standard system by means of 
measurements of 117 standard stars. The data were demonstrated to be on 
the standard system, and the limits in which these photometric results 
are valid were determined. Most target stars are indeed OB stars, and 
each cluster contains several stars that are located in the pulsational 
instability strips of main sequence B stars.

These measurements are required to determine the effective temperatures 
and luminosities of the targets. Published $uvby\beta$ photometry of the 
target clusters may now be tied into the standard system, allowing 
investigations of the clusters themselves, in terms of (differential) 
reddening, distance, etc. This is the foundation for several forthcoming 
papers devoted to individual clusters, including searches for stellar 
variability. Balona et al. (\cite{BPD11}) discuss the variability of the 
{\it Kepler} targets in detail.

\begin{acknowledgements}
This research is supported by the Austrian Fonds zur F\"orderung der 
wissenschaftlichen Forschung under grant P20526-N16. This research has 
made use of the WEBDA database, operated at the Institute for Astronomy 
of the University of Vienna.
\end{acknowledgements}

\clearpage
\onecolumn
\setcounter{table}{2}
\tablecaption{Str\"omgren-Crawford colour indices of standard stars.}
\begin{supertabular}{lcccccc}
\hline
Star & $N_{uvby}$ & $V$ & $(b-y)$ & $m_1$ & $c_1$ & $\beta$\\
\hline
BD$-$10 4682 & $1$ & $9.608*$ & $0.452$ & $-0.098$ & $-0.089$ & $2.602$\\
BD+38 4883   & $1$ & $9.471$ & $ 0.014$ & $ 0.108$ & $ 0.751$ & $2.788$\\
BD+39 4890   & $1$ & $9.468$ & $ 0.021$ & $ 0.122$ & $ 0.829$ & $2.850$\\
BD+39 4926   & $1$ & $9.269$ & $ 0.170$ & $ 0.044$ & $ 1.513$ & $2.312*$\\
BD+60 498    & $1$ & $9.938$ & $ 0.484$ & $-0.148*$ & $ 0.060$ & $2.615*$\\
BD+60 501    & $1$ & $9.597$ & $ 0.439$ & $-0.138$ & $-0.033$ & $2.598$\\
BD+61 2380   & $1$ & $9.141$ & $ 0.142$ & $ 0.081$ & $ 0.931$ & $2.836$\\
BD+62 2142   & $1$ & $9.033$ & $ 0.340$ & $-0.033$ & $ 0.259$ & $2.671$\\
BD+62 2150   & $1$ & $9.769$ & $ 0.381$ & $-0.043$ & $ 0.377$ & $2.706$\\
BD+62 2158   & $1$ & $10.092$& $ 0.194$ & $ 0.057$ & $ 0.937$ & $2.826$\\
BD+63 1907   & $1$ & $9.105$ & $ 0.680$ & $-0.133$ & $ 0.033$ & $2.548$\\
HD 13268     & $1$ & $8.155$ & $ 0.163$ & $-0.021$ & $-0.084$ & $2.573$\\
HD 14633     & $2$ & $7.442$ & $-0.080$ & $ 0.044$ & $-0.138$ & $2.553$\\
HD 15137     & $1$ & $7.852$ & $ 0.112$ & $-0.006$ & $-0.079$ & $2.563$\\
HD 161923    & $1$ & $9.135$ & $ 0.262$ & $ 0.105$ & $ 1.108$ & $2.879$\\
HD 175876    & $2$ & $6.939$ & $-0.018$ & $ 0.047$ & $-0.156$ & $2.572$\\
HD 179589    & $1$ & $9.159$ & $ 0.253$ & $ 0.175$ & $ 0.612$ & ...\\
HD 186980    & $2$ & $7.490$ & $ 0.131$ & $-0.002$ & $-0.108$ & $2.569$\\
HD 201345    & $1$ & $7.775*$ & $-0.027$ & $0.032$ & $-0.121$ & $ 2.567$\\
HD 207538    & $1$ & $7.310$ & $ 0.296$ & $-0.042$ & $-0.046$ & $2.597$\\
HD 210809    & $2$ & $7.588$ & $ 0.092$ & $ 0.027$ & $-0.122$ & $2.551$\\
HD 212883    & $3$ & $6.461$ & $-0.054$ & $ 0.081$ & $ 0.188$ & $2.651$\\
HD 213421    & $1$ & $8.243$ & $ 0.069$ & $ 0.163$ & $ 0.994$ & $2.875$\\
HD 213801    & $1$ & $8.164$ & $-0.011$ & $ 0.109$ & $ 0.625$ & $2.772$\\
HD 213976    & $1$ & $7.018$ & $-0.034$ & $ 0.069$ & $ 0.110$ & $2.640$\\
HD 214022    & $1$ & $8.518$ & $-0.006$ & $ 0.094$ & $ 0.466$ & $2.736$\\
HD 214168    & $1$ & $6.474$ & $-0.061$ & $ 0.080$ & $0.113*$ & $2.646$\\
HD 214180    & $1$ & $9.505$ & $ 0.065$ & $ 0.140$ & $ 1.051$ & $2.888$\\
HD 214243    & $1$ & $8.302$ & $-0.037$ & $ 0.090$ & $ 0.328$ & $2.700$\\
HD 214263    & $1$ & $6.826$ & $-0.041$ & $ 0.074$ & $ 0.148$ & $2.637$\\
HD 214432    & $1$ & $7.572$ & $-0.031$ & $ 0.088$ & $ 0.255$ & $2.673$\\
HD 214652    & $1$ & $6.871$ & $-0.028$ & $ 0.081$ & $ 0.183$ & $2.652$\\
HD 214783    & $1$ & $8.683$ & $ 0.041$ & $ 0.118$ & $ 1.086$ & $2.781$\\
HD 215191    & $1$ & $6.427$ & $-0.025$ & $ 0.067$ & $ 0.092$ & $2.626$\\
HD 215211    & $1$ & $8.656$ & $-0.013$ & $ 0.105$ & $ 0.547$ & $2.749$\\
HD 215212    & $1$ & $9.251$ & $ 0.037$ & $ 0.088$ & $ 0.648$ & ...\\
HD 216534    & $1$ & $8.524$ & $ 0.063$ & $ 0.047$ & $ 0.319$ & ...\\
HD 216684    & $1$ & $7.783$ & $ 0.051$ & $ 0.053$ & $ 0.313$ & ...\\
HD 216898    & $1$ & $8.018$ & $ 0.463$ & $-0.110$ & $ 0.010$ & $2.607$\\
HD 216926    & $1$ & $8.876$ & $ 0.200$ & $ 0.066$ & $ 0.886$ & $2.817$\\
HD 217086    & $1$ & $7.644$ & $ 0.536$ & $-0.138$ & $-0.004$ & $2.592$\\
HD 217101    & $1$ & $6.168$ & $-0.061$ & $ 0.086$ & $ 0.128$ & $2.645$\\
HD 218229    & $1$ & $8.163$ & $ 0.220$ & $-0.021$ & $ 0.860$ & $2.739$\\
HD 218407    & $1$ & $6.664$ & $ 0.008$ & $ 0.066$ & $ 0.200$ & $2.654$\\
HD 218450    & $1$ & $8.575$ & $ 0.072$ & $ 0.087$ & $ 0.915$ & $2.765$\\
HD 218915    & $1$ & $7.239$ & $ 0.063$ & $ 0.024$ & $-0.116$ & $2.553$\\
HD 227245    & $1$ & $9.754$ & $ 0.520$ & $-0.121$ & $-0.026$ & $2.590$\\
HD 235673    & $1$ & $9.150$ & $ 0.208$ & $-0.006$ & $-0.119$ & $2.575$\\
HR 6690      & $1$ & $6.287*$ & $0.019$ & $0.098$ & $0.799$ & ...\\
IC 4665 22   & $1$ & $8.722$ & $ 0.081$ & $ 0.084$ & $ 0.807$ & ...\\
NGC 869 3    & $1$ & $7.359$ & $ 0.234$ & $-0.058$ & $ 0.026$ & $2.575$\\
NGC 869 146  & $1$ & $9.174$ & $ 0.186$ & $-0.039$ & $ 0.061$ & $2.602$\\
NGC 869 339  & $1$ & $8.846$ & $ 0.288$ & $-0.087$ & $ 0.066$ & $2.602$\\
NGC 869 612  & $1$ & $8.440$ & $ 0.244$ & $-0.064$ & $ 0.055$ & $2.595$\\
NGC 869 662  & $1$ & $8.187$ & $ 0.293$ & $-0.084$ & $ 0.064$ & $2.588$\\
NGC 869 717  & $1$ & $9.264$ & $ 0.275$ & $-0.066$ & $ 0.091$ & $2.604$\\
NGC 869 782  & $1$ & $9.338*$ & $0.275$ & $-0.060$ & $0.170$ & $2.613$\\
NGC 869 847  & $1$ & $9.109$ & $ 0.343$ & $-0.097$ & $ 0.159$ & $2.592$\\
NGC 869 864  & $1$ & $  9.969$ & $0.283$ & $-0.059$ & $0.202$ & $2.630$\\
NGC 869 950  & $1$ & $ 11.281$ & $0.324$ & $-0.059$ & $0.218$ & $2.654$\\
NGC 869 978  & $1$ & $ 10.653$ & $0.310$ & $-0.052$ & $0.184$ & $2.630$\\
NGC 869 1004 & $1$ & $ 10.880$ & $0.301$ & $-0.046$ & $0.218$ & $2.640*$\\
NGC 869 1162 & $1$ & $  6.642$ & $0.473$ & $-0.117$ & $0.073$ & $2.555$\\
NGC 869 1187 & $1$ & $ 10.822$ & $0.378$ & $-0.072$ & $0.210$ & $2.639$\\
NGC 884 2139 & $1$ & $ 11.327$ & $0.302*$ & $-0.061$ & $0.194$ & $2.646$\\
NGC 884 2172 & $1$ & $  8.476$ & $0.211$ & $-0.026$ & $-0.107$ & $2.568$\\
NGC 884 2185 & $1$ & $ 10.942$ & $0.298$ & $-0.038$ & $0.385$ & $2.701$\\
NGC 884 2196 & $1$ & $ 11.544$ & $0.325*$ & $-0.067*$ & $0.216$ & ...\\
NGC 884 2227 & $3$ & $  8.045$ & $0.358$ & $-0.097$ & $0.110$ & $2.589$\\
NGC 884 2232 & $1$ & $ 11.047$ & $0.266$ & $-0.060*$ & $0.172$ & $2.631$\\
NGC 884 2246 & $1$ & $  9.915$ & $0.315$ & $-0.072$ & $0.113$ & $2.616$\\
NGC 884 2251 & $1$ & $ 11.549$ & $0.322$ & $-0.047$ & $0.361$ & $2.701$\\
NGC 884 2262 & $1$ & $ 10.559$ & $0.366$ & $-0.110$ & $0.179$ & $2.622$\\
NGC 884 2284 & $3$ & $  9.676$ & $0.367$ & $-0.130$ & $-0.017$& $2.416*$\\
NGC 884 2296 & $3$ & $  8.515$ & $0.289$ & $-0.082$ & $0.113$ & $2.591$\\
NGC 884 2299 & $1$ & $  9.130$ & $0.291$ & $-0.076$ & $0.123$ & $2.611$\\
NGC 884 2330 & $1$ & $ 11.442$ & $0.276$ & $-0.046$ & $0.242$ & ...\\
NGC 884 2572 & $1$ & $  9.998$ & $0.340$ & $-0.084$ & $0.174$ & $2.629$\\
NGC 884 2621 & $3$ & $  6.959$ & $0.523$ & $-0.122$ & $0.462$ & $2.590$\\
NGC 1502 1   & $1$ & $  6.944$ & $0.432$ & $-0.116$ & $0.027$ & $2.580$\\
NGC 1502 2   & $2$ & $7.100*$ & $0.399$ & $-0.105$ & $0.036$ & $2.592$\\
NGC 1502 16  & $1$ & $ 11.667$ & $0.493$ & $-0.048$ & $0.622$ & $2.749$\\
NGC 1502 26  & $1$ & $  9.651$ & $0.480$ & $-0.087$ & $0.161$ & $2.631$\\
NGC 1502 30  & $1$ & $  9.646$ & $0.481$ & $-0.083$ & $0.144$ & $2.629$\\
NGC 1502 35  & $1$ & $ 10.451$ & $0.428$ & $-0.048$ & $0.271$ & $2.659$\\
NGC 1502 36  & $1$ & $  9.790$ & $0.454$ & $-0.063$ & $0.196$ & $2.638$\\
NGC 1502 42  & $1$ & $ 12.593$ & $0.479$ & $-0.007*$ & $0.868*$ & $2.826*$\\
NGC 1502 43  & $1$ & $ 11.367$ & $0.455$ & $-0.051$ & $0.466$ & $2.707$\\
NGC 1502 45  & $1$ & $ 11.433$ & $0.473$ & $-0.046$ & $0.525$ & $2.722$\\
NGC 1502 52  & $1$ & $ 12.271$ & $0.530$ & $-0.047$ & $0.968$ & $2.827*$\\
NGC 1893 14  & $0$ &   ...   &  ...  &   ...  &  ...  & $ 2.596$\\
NGC 2169 17  & $1$ & $ 11.658$ & $0.154$ & $ 0.156$ & $1.051$ & $2.908$\\
NGC 2169 18  & $1$ & $ 11.821$ & $0.133$ & $ 0.102$ & $0.922$ & $2.867$\\
NGC 2244 114 & $2$ & $  7.631$ & $0.205$ & $-0.029$ & $-0.090$& ...\\
NGC 6871 6   & $1$ & $  8.729$ & $0.354$ & $-0.099$ & $-0.133*$ & ...\\
NGC 6871 7   & $1$ & $  8.788$ & $0.207$ & $-0.014$ & $0.044$ &  ... \\
NGC 6910 1   & $3$ & $  8.098$ & $0.090$ & $ 0.013$ & $0.082$ & $2.615$\\
NGC 6910 4   & $1$ & $  8.528$ & $0.704$ & $-0.149$ & $0.064$ & $2.583$\\
NGC 6910 5   & $1$ & $  9.664$ & $0.701$ & $-0.175$ & $0.127$ & $2.599$\\
NGC 6910 13  & $1$ & $ 10.307$ & $0.719$ & $-0.168$ & $0.169$ & $2.626$\\
NGC 6910 14  & $1$ & $ 10.349$ & $0.660$ & $-0.165$ & $0.115$ & $2.605$\\
NGC 6910 18  & $2$ & $ 10.776$ & $0.585$ & $-0.122$ & $0.160$ & $2.625$\\
NGC 6910 21  & $1$ & $ 11.756$ & $0.585$ & $-0.094$ & $0.208$ & $2.648$\\
NGC 6910 24  & $1$ & $ 11.706$ & $0.626$ & $-0.126$ & $0.217$ & $2.640$\\
NGC 6910 27  & $1$ & $ 11.680$ & $0.812$ & $-0.191$ & $0.198$ & $2.630$\\
NGC 6910 28  & $1$ & $ 12.241$ & $0.578$ & $-0.044*$ & $0.353$ & $2.668$\\
NGC 6913 1   & $1$ & $  8.871$ & $0.719$ & $-0.162$ & $0.160$ & $2.594$\\
NGC 6913 2   & $1$ & $  8.904$ & $0.620$ & $-0.132$ & $0.097$ & $2.597$\\
NGC 6913 3   & $2$ & $  8.966$ & $0.672$ & $-0.153$ & $0.136$ & $2.600$\\
NGC 6913 4   & $1$ & $ 10.190$ & $0.612$ & $-0.128$ & $0.158$ & $2.616$\\
NGC 6913 5   & $1$ & $  9.341$ & $0.627$ & $-0.131$ & $0.122$ & $2.595$\\
NGC 6913 7   & $2$ & $ 12.100$ & $0.691$ & $-0.116$ & $0.396*$ & ...\\
NGC 6913 9   & $1$ & $ 11.741$ & $0.601$ & $-0.099$ & $0.342$ & $2.652$\\
NGC 6913 27  & $1$ & $ 11.389$ & $0.666$ & $-0.101$ & $0.365$ & $2.659$\\
NGC 6913 63  & $1$ & $ 10.543$ & $0.176$ & $ 0.027$ & $0.485$ & $2.726$\\
NGC 6913 64  & $1$ & $ 10.099$ & $0.157$ & $ 0.021$ & $0.477$ & $2.739$\\
NGC 7380 2   & $2$ & $  8.546$ & $0.329$ & $-0.059$ & $-0.067$& $2.585$\\
\hline
\end{supertabular}
\tablefoot{$N_{uvby}$ is the number of $uvby$ measurements 
per star. Each star was only measured once in $H_{\beta}$ except HD 
161923 that was measured twice. Values marked with asterisks were excluded 
from the determinations of the coefficients for the transformation equations.}

\clearpage

\setcounter{table}{3}
\tablecaption{Str\"omgren-Crawford colour indices of open cluster target 
stars.}
\begin{supertabular}{lccccccc}
\hline
Star & $N_{uvby}$ & $V$ & $(b-y)$ & $m_1$ & $c_1$ & $\beta$ & $N_{\beta}$\\
\hline
ASCC 130 1      & $1$ & $10.670$ & $0.230$ & $0.000$  & $0.639$ & $2.711$ & $1$\\
ASCC 130 2      & $1$ & $11.618$ & $0.406$ & $-0.080$ & $0.078$ & $2.613$ & $1$\\
ASCC 130 4      & $1$ & $11.116$ & $0.282$ & $-0.025$ & $0.446$ & $2.769$ & $1$\\
ASCC 130 5      & $1$ & $10.160$ & $0.241$ & $0.001$  & $0.600$ & $2.707$ & $1$\\
ASCC 130 6      & $1$ & $11.380$ & $0.267$ & $0.002$  & $0.345$ & $2.666$ & $1$\\
ASCC 130 8      & $1$ & $11.243$ & $0.335$ & $-0.039$ & $0.203$ & $2.657$ & $1$\\
ASCC 130 19     & $1$ & $ 9.690$ & $0.271$ & $-0.054$ & $-0.026$& $2.597$ & $1$\\
Berkeley 4 9    & $1$ & $10.923$ & $0.393$ & $0.203:$ & $0.329$ &  ...  & $0$\\
Berkeley 4 77   & $1$ & $11.162$ & $0.298$ & $0.139$  & $0.885$ &  ...  & $0$\\
Berkeley 4 101  & $1$ & $12.300$ & $0.470$ & $-0.110$ & $0.263$ & $2.652$ & $1$\\
Berkeley 4 115  & $1$ & $11.858$ & $0.323$ & $0.143$  & $0.864$ & $2.839$ & $1$\\
Berkeley 4 210  & $1$ & $11.490$ & $0.446$ & $-0.075$ & $0.246$ & $2.612$ & $1$\\
Berkeley 4 238  & $1$ & $12.294$ & $0.451$ & $-0.060$ & $0.357$ & $2.723$ & $1$\\
Berkeley 4 513  & $1$ & $11.996$ & $0.489$ & $-0.124$ & $0.396$ & $2.577$ & $1$\\
Berkeley 4 649  & $1$ & $12.165$ & $0.358$ & $-0.034$ & $0.290$ & $2.666$ & $1$\\
Berkeley 4 703  & $1$ & $10.633$ & $0.413$ & $-0.084$ & $0.044$ & $2.610$ & $1$\\
Berkeley 4 709  & $1$ & $11.929$ & $0.493$ & $0.152$  & $0.401$ &  ...  & $0$\\
Berkeley 4 794  & $1$ & $11.633$ & $0.504$ & $-0.098$ & $0.065$ & $2.629$ & $1$\\
Berkeley 4 877  & $1$ & $12.193$ & $0.407$ & $-0.061$ & $0.239$ & $2.654$ & $1$\\
Berkeley 4 966  & $1$ & $10.528$ & $0.123$ & $0.113$  & $1.009$ & $2.901$ & $1$\\
Berkeley 4 977  & $1$ & $10.771$ & $0.446$ & $0.247:$ & $0.370$ &  ...  & $0$\\
Berkeley 4 1008 & $1$ & $10.798$ & $0.222$ & $0.019$  & $0.822$ &  ...  & $0$\\
Berkeley 4 1084 & $1$ & $10.662$ & $0.458$ & $-0.112$ & $0.175$ & $2.622$ & $1$\\
Berkeley 4 1101 & $1$ & $11.086$ & $0.254$ & $0.151$  & $0.806$ &  ...  & $0$\\
Berkeley 4 1110 & $1$ & $10.800$ & $0.463$ & $0.228:$ & $0.342$ &  ...  & $0$\\
Berkeley 4 1142 & $1$ & $11.386$ & $0.343$ & $0.119$  & $0.916$ &  ...  & $0$\\
Berkeley 4 1204 & $1$ & $11.622$ & $0.443$ & $-0.086$ & $0.157$ & $2.635$ & $1$\\
Berkeley 4 1253 & $1$ & $12.109$ & $0.329$ & $0.115$  & $0.938$ &  ...  & $0$\\
Berkeley 4 1302 & $1$ & $12.038$ & $0.455$ & $-0.096$ & $0.019$ & $2.621$ & $1$\\
Berkeley 4 1317 & $1$ & $10.483$ & $0.440$ & $-0.108$ & $0.183$ & $2.637$ & $1$\\
Berkeley 4 1327 & $1$ & $12.262$ & $0.401$ & $0.190:$ & $0.327$ & $2.652$ & $1$\\
Berkeley 4 1356 & $1$ & $11.809$ & $0.445$ & $0.115$  & $0.466$ &  ...  & $0$\\
Berkeley 4 1386 & $1$ & $11.724$ & $0.547$ & $-0.132$ & $0.232$ & $2.653$ & $2$\\
Berkeley 4 2000 & $1$ & $10.054$ & $0.329$ & $0.141$  & $0.768$ & $2.754$ & $1$\\
Berkeley 4 2001 & $1$ & $ 9.845$ & $0.447$ & $-0.123$ & $0.018$ & $2.591$ & $1$\\
Berkeley 4 2002 & $2$ & $11.444$ & $0.455$ & $-0.094$ & $0.106$ & $2.638$ & $1$\\
Berkeley 4 2003 & $3$ & $ 9.486$ & $0.537$ & $-0.130$ & $0.042$ & $2.580$ & $1$\\
Berkeley 4 2005 & $1$ & $11.047$ & $0.468$ & $-0.098$ & $0.291$ & $2.641$ & $1$\\
Berkeley 4 2007 & $1$ & $12.128$ & $0.346$ & $0.037$  & $1.027$ & $2.913$ & $1$\\
NGC 637 1       & $1$ & $ 9.979$ & $0.358$ & $-0.075$ & $0.090$ & $2.604$ & $1$\\
NGC 637 3       & $1$ & $10.578$ & $0.366$ & $-0.072$ & $0.185$ & $2.640$ & $1$\\
NGC 637 6       & $1$ & $10.351$ & $0.373$ & $-0.089$ & $0.092$ & $2.613$ & $1$\\
NGC 637 7       & $1$ & $10.670$ & $0.360$ & $-0.076$ & $0.129$ & $2.623$ & $1$\\
NGC 637 137     & $1$ & $10.787$ & $0.349$ & $-0.058$ & $0.190$ & $2.653$ & $1$\\
NGC 637 138     & $1$ & $10.158$ & $0.327$ & $-0.075$ & $0.106$ & $2.619$ & $1$\\
NGC 1893 13     & $1$ & $12.463$ & $0.287$ & $-0.004$ & $0.328$ & $2.684$ & $1$\\
NGC 1893 33     & $1$ & $12.287$ & $0.277$ & $-0.026$ & $0.143$ & $2.640$ & $1$\\
NGC 1893 59     & $1$ & $12.153$ & $0.225$ & $0.030$  & $0.125$ & $2.655$ & $1$\\
NGC 1893 106    & $1$ & $12.395$ & $0.303$ & $-0.019$ & $0.123$ & $2.656$ & $1$\\
NGC 1893 139    & $1$ & $12.023$ & $0.157$ & $0.055$  & $0.018$ & $2.622$ & $1$\\
NGC 1893 140    & $1$ & $12.415$ & $0.191$ & $0.051$  & $0.176$ & $2.668$ & $1$\\
NGC 1893 141    & $0$ &   ...  &  ...  &  ...   &  ...  & $2.620$ & $1$\\
NGC 1893 168    & $1$ & $12.329$ & $0.301$ & $-0.022$ & $0.190$ & $2.663$ & $1$\\
NGC 1893 196    & $1$ & $12.637$ & $0.420$ & $-0.050$ & $0.104$ & $2.441:$& $1$\\
NGC 1893 228    & $1$ & $12.525$ & $0.207$ & $0.044$  & $0.171$ & $2.659$ & $1$\\
NGC 1893 256    & $1$ & $11.870$ & $0.228$ & $0.032$  & $0.283$ & $2.629$ & $1$\\
NGC 1893 290    & $0$ &   ...  &  ...  &  ...   &  ...  & $2.612$ & $1$\\
NGC 1893 343    & $1$ & $10.916$ & $0.271$ & $-0.011$ & $0.008$ & $2.617$ & $1$\\
NGC 1893 345    & $1$ & $10.831$ & $0.282$ & $-0.015$ & $-0.004$ & $2.606$ & $1$\\
NGC 2244 172    & $1$ & $11.233$ & $0.274$ & $0.006$ & $0.245$ & $2.671$ & $1$\\
NGC 2244 190    & $1$ & $11.274$ & $0.245$ & $0.015$  & $0.278$ & $2.679$ & $1$\\
NGC 2244 193    & $0$ &   ...  &  ...  &  ...   &  ...  & $2.672$ & $1$\\
NGC 2244 239    & $1$ & $11.120$ & $0.254$ & $0.035$  & $0.474$ & $2.737$ & $1$\\
NGC 2244 241    & $1$ & $11.102$ & $0.236$ & $0.051$  & $0.491$ & $2.751$ & $1$\\
NGC 2244 279    & $1$ & $11.305$ & $0.308$ & $-0.020$ & $0.145$ & $2.481:$& $1$\\
NGC 2244 280    & $1$ & $10.881$ & $0.289$ & $0.000$ & $0.312$ & $2.674$ & $1$\\
NGC 2244 392    & $0$ &   ...  &  ...  &  ...   &  ...  & $2.694$ & $1$\\
NGC 2244 1034   & $1$ & $11.304$ & $0.320$ & $0.030$  & $0.540$ & $2.748$ & $1$\\
NGC 2244 1618   & $1$ & $10.972$ & $0.351$ & $0.186:$ & $0.333$ & $2.663$ & $1$\\
NGC 2244 3010   & $1$ & $10.925$ & $0.179$ & $0.052$  & $0.478$ & $2.731$ & $1$\\
NGC 7380 4      & $1$ & $10.194$ & $0.358$ & $-0.027$ & $0.571$ & $2.602$ & $1$\\
NGC 7380 7      & $1$ & $10.665$ & $0.373$ & $0.171:$ & $0.416$ &  ...  & $0$\\
NGC 7380 8      & $1$ & $10.641$ & $0.288$ & $-0.047$ & $0.052$ & $2.620$ & $1$\\
NGC 7380 9      & $1$ & $10.688$ & $0.342$ & $-0.044$ & $0.053$ & $2.625$ & $1$\\
NGC 7380 31     & $1$ & $10.617$ & $0.387$ & $-0.081$ & $-0.062$& $2.582$ & $1$\\
NGC 7380 34     & $1$ & $11.832$ & $0.333$ & $-0.024$ & $0.138$ & $2.642$ & $1$\\
NGC 7380 35     & $1$ & $11.869$ & $0.308$ & $-0.033$ & $0.146$ & $2.619$ & $1$\\
NGC 7380 36     & $1$ & $11.827$ & $0.296$ & $0.020$  & $0.960$ & $2.834$ & $1$\\
NGC 7380 37     & $1$ & $11.963$ & $0.319$ & $-0.046$ & $0.260$ & $2.658$ & $1$\\
NGC 7380 40     & $1$ & $12.158$ & $0.340$ & $-0.016$ & $0.374$ & $2.679$ & $1$\\
NGC 7380 41     & $1$ & $12.198$ & $0.335$ & $-0.023$ & $0.153$ & $2.642$ & $1$\\
NGC 7380 42     & $1$ & $12.290$ & $0.529$ & $-0.096$ & $0.135$ & $2.595$ & $1$\\
NGC 7380 134    & $1$ & $ 9.172$ & $0.401$ & $-0.089$ & $-0.042$& $2.581$ & $1$\\
NGC 7380 135    & $1$ & $10.347$ & $0.336$ & $-0.075$ & $0.016$ & $2.609$ & $1$\\
NGC 7380 136    & $1$ & $10.409$ & $0.447$ & $-0.079$ & $0.081$ & $2.615$ & $1$\\
NGC 7380 138    & $1$ & $11.216$ & $0.332$ & $-0.038$ & $0.083$ & $2.644$ & $1$\\
NGC 7380 184    & $1$ & $11.020$ & $0.168$ & $0.178:$ & $0.957$ &  ...  & $0$\\
NGC 7380 5476   & $1$ & $11.153$ & $0.377$ & $0.186:$ & $0.261$ &  ...  & $0$\\
NGC 7380 5593   & $1$ & $10.807$ & $0.226$ & $0.175:$ & $0.773$ & $2.815$ & $1$\\
NGC 7380 5596   & $1$ & $10.092$ & $0.302$ & $0.135$  & $0.441$ &  ...  & $0$\\
NGC 7380 5666   & $1$ & $11.260$ & $0.277$ & $-0.007$ & $0.649$ & $2.724$ & $1$\\
NGC 7380 5678   & $1$ & $11.344$ & $0.265$ & $0.049$  & $1.185$ & $2.860$ & $1$\\
NGC 7380 5681   & $1$ & $ 9.825$ & $0.234$ & $0.004$  & $0.731$ & $2.747$ & $1$\\
NGC 7380 5755   & $1$ & $10.616$ & $0.244$ & $-0.047$ & $0.064$ & $2.621$ & $1$\\
NGC 7380 5759   & $1$ & $10.489$ & $0.229$ & $0.021$  & $0.506$ & $2.715$ & $1$\\
NGC 7380 5761   & $1$ & $10.980$ & $0.237$ & $0.019$  & $0.602$ & $2.761$ & $1$\\
NGC 7380 5804   & $1$ & $10.614$ & $0.270$ & $0.146$  & $0.718$ &  ...  & $0$\\
Roslund 2 2     & $1$ & $10.495$ & $0.648$ & $-0.155$ & $0.084$ & $2.611$ & $1$\\
Roslund 2 6     & $1$ & $10.773$ & $0.239$ & $0.095$  & $1.074$ & $2.868$ & $1$\\
Roslund 2 7     & $1$ & $10.671$ & $0.566$ & $-0.102$ & $0.097$ & $2.587$ & $1$\\
Roslund 2 11    & $1$ & $ 8.749$ & $0.624$ & $-0.153$ & $-0.021$& $2.578$ & $1$\\
Roslund 2 13    & $1$ & $11.383$ & $1.201:$& $ 0.393:$& $0.536:$&  ...  & $0$\\
Roslund 2 14    & $1$ & $11.428$ & $0.448$ & $-0.018$ & $0.903$ & $2.843$ & $1$\\
Roslund 2 16    & $1$ & $ 9.274$ & $0.626$ & $-0.169$ & $0.082$ & $2.598$ & $1$\\
Roslund 2 17    & $1$ & $11.109$ & $0.570$ & $-0.134$ & $0.171$ & $2.639$ & $1$\\
Roslund 2 18    & $1$ & $ 7.843$ & $0.645$ & $-0.169$ & $0.034$ & $2.566$ & $1$\\
Roslund 2 21    & $1$ & $12.039$ & $0.778$ & $-0.150$ & $0.187$ &  ...  & $0$\\
\hline
\end{supertabular}
\tablefoot{$N_{uvby}$ is the number of $uvby$ measurements per 
star; $N_{\beta}$ is the number of $H_{\beta}$ observations. Entries 
marked with colons are outside the validity of the transformation 
equations.}

\clearpage

\setcounter{table}{4}
\tablecaption{Str\"omgren-Crawford photometry of known $\beta$~Cephei 
and other target stars}
\begin{supertabular}{lccccccc}
\hline
Star & $N_{uvby}$ & $V$ & $(b-y)$ & $m_1$ & $c_1$ & $\beta$ & $N_{\beta}$\\
\hline
BD+36 4867      & $1$ & $10.369$ & $ 0.579$ & $0.281:$ & $0.429$  & $2.609$ & $1$\\
GSC 03142-00038 & $1$ & $12.546$ & $ 0.265$ & $0.177:$ & $0.547$  & $2.719$ & $1$\\
GSC 06272-01557 & $1$ & $10.739$ & $ 0.656$ & $-0.142$ & $0.176$  & $2.621$ & $1$\\
HD 166540       & $1$ & $ 8.126$ & $ 0.197$ & $-0.013$ & $-0.029$ & $2.603$ & $2$\\
HD 167743       & $1$ & $ 9.650$ & $ 0.329$ & $-0.034$ & $0.095$  & $2.634$ & $1$\\
HD 180642       & $1$ & $ 8.221$ & $ 0.238$ & $-0.043$ & $0.009$  & $2.601$ & $1$\\
HD 203664       & $1$ & $ 8.512$ & $-0.086$ & $ 0.040$ & $-0.087$ & $2.572$ & $1$\\
HN Aqr          & $1$ & $11.408$ & $-0.085$ & $ 0.068$ & $0.043$  & $2.606$ & $1$\\
NGC 637 4       & $1$ & $10.782$ & $ 0.393$ & $-0.075$ & $0.150$  & $2.620$ & $1$\\
NGC 869 692     & $1$ & $ 9.369$ & $ 0.265$ & $-0.094$ & $0.044$  & $2.600$ & $1$\\
NGC 869 839     & $1$ & $ 9.371$ & $ 0.328$ & $-0.078$ & $0.109$  & $2.610$ & $1$\\
NGC 869 992     & $1$ & $10.016$ & $ 0.293$ & $-0.067$ & $0.241$  & $2.623$ & $1$\\
NGC 884 2085    & $1$ & $11.321$ & $ 0.290$ & $-0.044$ & $0.230$  & $2.624$ & $1$\\
NGC 884 2444    & $1$ & $ 9.503$ & $ 0.344$ & $-0.104$ & $0.140$  & $2.617$ & $1$\\
NGC 884 2566    & $1$ & $10.548$ & $ 0.439$ & $-0.106$ & $0.051$  & $2.528:$& $1$\\
NGC 6910 16     & $1$ & $10.439$ & $ 0.673$ & $-0.151$ & $0.173$  & $2.624$ & $1$\\
NGC 6910 25     & $1$ & $11.459$ & $ 0.763$ & $-0.180$ & $0.280$  & $2.637$ & $1$\\
NGC 7235 8      & $1$ & $11.906$ & $ 0.526$ & $-0.126$ & $0.139$  & $2.614$ & $1$\\
V909 Cas        & $1$ & $10.623$ & $ 0.370$ & $-0.072$ & $0.101$  & $2.619$ & $1$\\
\hline
\end{supertabular}
\tablefoot{$N_{uvby}$ is the number of $uvby$ measurements per star; 
$N_{\beta}$ is the number of $H_{\beta}$ observations. Entries marked 
with colons are outside the validity of the transformation equations.}

\clearpage

\setcounter{table}{5}
\tablecaption{Str\"omgren-Crawford colour indices of early-type stars in 
the {\it Kepler} field}
\begin{supertabular}{lccccccc}
\hline
Star & $N_{uvby}$ & $V$ & $(b-y)$ & $m_1$ & $c_1$ & $\beta$& $N_{\beta}$\\
\hline
KIC 3240411  & $1$ & $10.271$ & $-0.041$ & $0.079$ & $0.161$ & $2.643$ & $1$\\
KIC 3756031  & $1$ & $10.012$ & $-0.005$ & $0.084$ & $0.372$ & $2.696$ & $1$\\
KIC 3839930  & $1$ & $10.702$ & $-0.013$ & $0.087$ & $0.321$ & $2.709$ & $1$\\
KIC 3848385  & $1$ & $ 8.909$ & $0.043$  & $0.082$ & $0.785$ & $2.739$ & $1$\\
KIC 3865742  & $1$ & $11.120$ & $0.028$  & $0.072$ & $0.201$ & $2.662$ & $1$\\
KIC 4276892  & $1$ & $ 9.168$ & $0.026$  & $0.124$ & $1.062$ & $2.855$ & $1$\\
KIC 4581434  & $1$ & $ 9.111$ & $0.050$  & $0.159$ & $1.095$ & $2.896$ & $1$\\
KIC 4909697  & $1$ & $10.703$ & $0.230$  & $0.179:$& $0.987$ & $2.852$ & $1$\\
KIC 5130305  & $1$ & $10.143$ & $0.040$  & $0.112$ & $0.931$ & $2.843$ & $1$\\
KIC 5217845  & $1$ & $ 9.420$ & $0.081$  & $0.081$ & $0.784$ & $2.738$ & $1$\\
KIC 5304891  & $1$ & $ 9.163$ & $0.051$  & $0.085$ & $0.804$ & $2.747$ & $1$\\
KIC 5458880  & $4$ & $ 7.762$ & $0.010$  & $0.031$ & $-0.039$& $2.582$ & $1$\\
KIC 5479821  & $1$ & $ 9.803$ & $0.041$  & $0.083$ & $0.313$ & $2.699$ & $2$\\
KIC 5786771  & $1$ & $ 9.075$ & $-0.007$ & $0.152$ & $1.006$ & $2.867$ & $1$\\
KIC 6848529  & $1$ & $10.628$ & $-0.100$ & $0.096$ & $-0.029$& $2.645$ & $1$\\
KIC 7548479  & $1$ & $ 8.387$ & $0.141$  & $0.219:$& $0.774$ & $2.824$ & $1$\\
KIC 7599132  & $1$ & $ 9.333$ & $0.001$  & $0.123$ & $0.870$ & $2.830$ & $1$\\
KIC 7974841  & $3$ & $ 8.167$ & $0.026$  & $0.131$ & $0.838$ & $2.823$ & $2$\\
KIC 8018827  & $1$ & $ 8.020$ & $0.004$  & $0.136$ & $0.895$ & $2.833$ & $1$\\
KIC 8057661  & $1$ & $11.613$ & $0.207$  & $0.002$ & $0.196$ & $2.656$ & $1$\\
KIC 8161798  & $1$ & $10.396$ & $-0.002$ & $0.191:$& $0.630$ & $2.793$ & $1$\\
KIC 8177087  & $1$ & $ 8.108$ & $0.008$  & $0.080$ & $0.589$ & $2.707$ & $1$\\
KIC 8324268  & $1$ & $ 7.922$ & $-0.019$ & $0.144$ & $0.488$ & $2.744$ & $1$\\
KIC 8351193  & $1$ & $ 7.580$ & $-0.029$ & $0.145$ & $0.880$ & $2.863$ & $1$\\
KIC 8381949  & $1$ & $11.010$ & $0.073$  & $0.041$ & $0.157$ & $2.633$ & $1$\\
KIC 8389948  & $1$ & $ 9.206$ & $0.073$  & $0.121$ & $1.023$ & $2.858$ & $1$\\
KIC 8415752  & $1$ & $10.598$ & $0.103$  & $0.219:$& $0.867$ & $2.845$ & $1$\\
KIC 8459899  & $1$ & $ 8.674$ & $0.021$  & $0.075$ & $0.398$ & $2.693$ & $1$\\
KIC 8488717  & $1$ & $11.658$ & $0.013$  & $0.146$ & $0.921$ & $2.849$ & $1$\\
KIC 8692626  & $1$ & $ 8.308$ & $0.058$  & $0.241:$& $0.933$ & $2.889$ & $1$\\
KIC 8714886  & $1$ & $10.866$ & $0.118$  & $0.060$ & $0.286$ & $2.694$ & $1$\\
KIC 8766405  & $1$ & $ 8.825$ & $0.001$  & $0.081$ & $0.525$ & $2.692$ & $1$\\
KIC 9964614  & $1$ & $10.683$ & $-0.013$ & $0.063$ & $0.193$ & $2.638$ & $1$\\
KIC 10130954 & $1$ & $11.015$ & $-0.047$ & $0.076$ & $0.238$ & $2.658$ & $1$\\
KIC 10285114 & $1$ & $11.121$ & $-0.035$ & $0.088$ & $0.371$ & $2.695$ & $1$\\
KIC 10797526 & $1$ & $ 8.300$ & $-0.028$ & $0.060$ & $0.061$ & $2.598$ & $1$\\
KIC 10960750 & $1$ & $ 9.833$ & $-0.065$ & $0.080$ & $0.165$ & $2.640$ & $1$\\
KIC 11360704 & $1$ & $10.650$ & $-0.032$ & $0.081$ & $0.286$ & $2.662$ & $1$\\
KIC 11817929 & $1$ & $10.301$ & $-0.058$ & $0.119$ & $0.637$ & $2.738$ & $1$\\
KIC 11973705 & $1$ & $ 9.074$ & $0.148$  & $0.152$ & $0.750$ & $2.777$ & $1$\\
KIC 12217324 & $2$ & $ 8.267$ & $-0.038$ & $0.149$ & $0.937$ & $2.828$ & $2$\\
KIC 12258330 & $1$ & $ 9.402$ & $-0.050$ & $0.099$ & $0.355$ & $2.706$ & $1$\\
\hline
\end{supertabular}
\tablefoot{$N_{uvby}$ is the number of $uvby$ measurements per star; 
$N_{\beta}$ is the number of observations in $H_{\beta}$. Entries marked 
with colons are outside the validity of the transformation equations.}

\clearpage

\begin{thebibliography}{}
%
\bibitem[2011]{BPD11}Balona, L. A., Pigulski, A., De Cat, P., et al., 
2011, MNRAS, in press
\bibitem[2010]{BBK10}Batalha, N. M., Borucki, W. J., Koch, D. G., et al., 
2010, ApJ, 713, L109
\bibitem[1969]{B69}Bevington, P. R., {\it Data reduction and error 
analysis for the physical sciences}, McGraw-Hill, New York, 1969, p.~3
\bibitem[1975]{Cr75}Crawford, D. L., 1975, PASP, 87, 481
\bibitem[1994]{Cr94}Crawford, D. L., 1994, PASP, 106, 397
\bibitem[1994]{Cr99}Crawford, D. L., 1999, in {\it CCD Precision 
Photometry Workshop}, ed. R. Craine et al., ASP Conf. Ser., 189, 6
\bibitem[1970]{CB70}Crawford, D. L., \& Barnes, J. V., 1970, AJ, 75, 952
\bibitem[1972]{CB72}Crawford, D. L., \& Barnes, J. V., 1972, AJ, 77, 862
\bibitem[1976]{CW76}Crawford, D. L., \& Warren, W. H., 1976, PASP, 88, 930
\bibitem[1971]{CBG71}Crawford, D. L., Barnes, J. V., \& Golson, J. C., 
1971, AJ, 76, 1058
\bibitem[1977]{CBH77}Crawford, D. L., Barnes, J. V., \& Hill, G., 1977, 
AJ, 82, 606
\bibitem[1970]{CGP70}Crawford, D. L., Glaspey, J. W., \& Perry, C. L., 
1970, AJ, 75, 822
\bibitem[1972]{CBG72}Crawford, D. L., Barnes, J. V., Gibson, J., et al., 
1972, A\&AS, 5, 109
\bibitem[2007]{PDC07}De Cat, P., 2007, CoAst, 150, 167
%\bibitem[2009]{WD09}Dziembowski, W. A., 2009, CoAst, 158, 227
\bibitem[1996]{F96}Flower, P. J., 1996, ApJ, 469, 355
\bibitem[1959]{HRS59}Hardorp, J., Rohlfs, K., Slettebak, A., \& 
Stock, J., 1959, Publ. Hamburger Sternw., Warner \& Swasey Obs., 1
\bibitem[1965]{HA65}Hoag, A. A., \& Applequist, N. L., 1965, ApJS, 12, 215
\bibitem[1977]{JK77}Knude, J. K., 1977, A\&AS, 30, 297
\bibitem[2010]{KBB10}Koch, D. G., Borucki, W. J., Basri, G., et al., 2010, 
ApJ, 713, L79
\bibitem[2000]{L00}Livingston, W. C., 2000, in {\it Allen's 
Astrophysical Quantities}, 4$^{\rm th}$ edition, ed. A. N. Cox, Springer
Verlag, p.~341
\bibitem[2001]{MBN01}Marco, A., Bernabeu, G., \& Negueruela, I., 2001, AJ, 121, 2075
\bibitem[2009]{MHG09}McSwain, M. V., Huang, W., \& Gies, D. R., 2009, 
ApJ, 700, 1216
\bibitem[1997]{MMH97}Mermilliod, J.-C., Mermilliod, M., \& Hauck, B., 
1997, A\&AS, 124, 349
\bibitem[1992]{MD92}Moskalik, P., \& Dziembowski, W. A., 1992, A\&A, 256, L5
\bibitem[1993]{NSW93}Napiwotzki, R., Sch\"onberner, D., \& Wenske, V., 
1993, A\&A, 268, 653
\bibitem[1972]{PLB78}Perry, C. L., Lee, P. D., \& Barnes, J. V., 1978, 
PASP, 90, 73
\bibitem[2008]{PP08}Pigulski, A., \& Pojma{\'n}ski, G., 2008, A\&A 477, 
917
\bibitem[1994]{RI94}Rogers, F. J., \& Iglesias, C. A., 1994, Science, 263, 50
\bibitem[2005]{SH05}Stankov, A., \& Handler, G., 2005, ApJS, 158, 193
\bibitem[2008]{ZP08}Zdravkov, T., \& Pamyatnykh, A. A., 2008, JPhCS 118, 
012079
\end{thebibliography}
\end{document}